\documentclass[preprint]{aastex}

\slugcomment{}

\shortauthors{Matheson et al.}
\shorttitle{Supernova 1993J}

\begin{document}


\title{Optical Spectroscopy of Supernova 1993J During Its First 2500 Days}

\author{Thomas Matheson\altaffilmark{1}, Alexei
V. Filippenko\altaffilmark{1}, Aaron J. Barth\altaffilmark{2}, Luis
C. Ho\altaffilmark{3}, Douglas C. Leonard\altaffilmark{1},
Matthew A. Bershady\altaffilmark{4}, Marc
Davis\altaffilmark{1}, David S.  Finley\altaffilmark{5}, David
Fisher\altaffilmark{6,7}, Rosa A.  Gonz\'alez\altaffilmark{8}, Suzanne
L. Hawley\altaffilmark{9}, David C. Koo\altaffilmark{6}, Weidong
Li\altaffilmark{1}, Carol J. Lonsdale\altaffilmark{10}, David
Schlegel\altaffilmark{11}, Harding E. Smith\altaffilmark{12}, Hyron
Spinrad\altaffilmark{1}, 
 and Gregory D. Wirth\altaffilmark{13}}

\altaffiltext{1}{Department of Astronomy, University of California,
    Berkeley, CA 94720-3411}
\altaffiltext{2}{Harvard-Smithsonian Center for Astrophysics,
    60 Garden Street, Cambridge, MA 02138}
\altaffiltext{3}{The Observatories of the Carnegie Institution of
    Washington, 813 Santa Barbara Street, Pasadena, CA 91101-1292}
\altaffiltext{4}{Department of Astronomy, University of Wisconsin, 475
    North Charter Street, Madison, WI 53706}
\altaffiltext{5}{Eureka Scientific, Inc., 2452 Delmer Street, Suite
    100, Oakland, CA 94602}
\altaffiltext{6}{University of California Observatories/Lick
 Observatory, Department of Astronomy and Astrophysics, University of
 California, Santa Cruz, CA 95064}
\altaffiltext{7}{Current Address: U.S. Department of State,
    U.S. Consulate General, Ho Chi Minh City, Vietnam}
\altaffiltext{8}{Observatoire de Gen\`eve, Mailletes 51, 1290
    Sauverny, Switzerland}
\altaffiltext{9}{Astronomy Department, University of Washington, Box
    351580, Seattle, WA 98195-1580 }
\altaffiltext{10}{Infrared Processing and Analysis Center, California
    Institute of Technology 100-22, Pasadena, CA 91125} 
\altaffiltext{11}{Department of Astrophysical Sciences, Princeton University,
    Peyton Hall, Princeton, NJ 08544} 
\altaffiltext{12}{Center for Astrophysics and Space Sciences,
    University of California, San Diego, La Jolla, CA 92093-0424} 
\altaffiltext{13}{W. M. Keck Observatory, Kamuela, HI 96743}


\begin{abstract}
We present 42 low-resolution spectra of Supernova (SN) 1993J, our
complete collection from the Lick and Keck Observatories, from day 3
after explosion to day 2454, as well as one Keck high-dispersion
spectrum from day 383.  SN 1993J began as an apparent SN II, albeit an
unusual one.  After a few weeks, a dramatic transition took place, as
prominent helium lines emerged in the spectrum.  SN 1993J had
metamorphosed from a SN II to a SN IIb.  Nebular spectra of SN 1993J
closely resemble those of SNe Ib and Ic, but with a persistent
H$\alpha$ line.  At very late times, the H$\alpha$ emission line
dominated the spectrum, but with an unusual, box-like profile.  This
is interpreted as an indication of circumstellar interaction.

\end{abstract}



\section{Introduction}

Supernova 1993J was visually discovered in the nearby galaxy M81 (NGC
3031; $d$ = 3.6 Mpc; Freedman et al. 1994) by F. Garcia on 1993 March
28.906 UT (Ripero, Garcia, \& Rodriguez 1993; note that all calendar
dates used herein are UT).  It reached a maximum brightness of $m_V =
10.8$ mag (e.g., Richmond et al. 1994), becoming the brightest supernova
(SN) in the northern hemisphere since SN 1954A ($m_{\rm{pg}} = 9.95$;
Wild 1960; Barbon, Ciatti, \& Rosino 1973).  In terms of observational
coverage, both in temporal consistency (almost nightly observations
at early times) and in the details of individual observations
(including observations with signal-to-noise [S/N] ratios, spectral
resolutions, and wavelength regions not typically found in studies of
SNe), SN 1993J is surpassed only by SN 1987A.  Early spectra showed an
almost featureless blue continuum, possibly with broad, but weak,
H$\alpha$ and \ion{He}{1} $\lambda$5876 lines.  This led to a
classification of the SN as Type II (Filippenko et al. 1993; Garnavich
\& Ann 1993; see Filippenko 1997 for a general discussion of SN
types).  Wheeler \& Filippenko (1996) present a thorough review of the
early work on SN 1993J.

Both the spectra and the light curve of SN 1993J quickly began to
indicate that this was not a typical Type II SN.  Indeed, the
initially unusual light curve and the appearance of \ion{He}{1} lines
in the spectra was interpreted as evidence that SN 1993J was similar
to a SN Ib, with a low-mass outer layer of hydrogen that gave the
early impression of a SN II (see discussion and references below).
Following Woosley et al. (1987), it was described as a ``Type IIb''
SN.  This transformation from SN II to nearly SN Ib indicates a common
mechanism (core collapse) for these two observationally defined
subclasses.  SN 1993J is thus one of the most significant SNe ever
studied, not only for its role in linking Types II and Ib (and
possibly Ic), but also because it was observed with such great detail.

In \S 2 of this paper we review the study of SN 1993J; we are
presenting over six years of optical spectra and it is appropriate to
provide such a summary.  (Wheeler \& Filippenko [1996] only cover the
first few months of the development of SN 1993J in their review;
moreover, the interpretation to follow depends on the context of the
previous observations and analyses.)  We then present the collection
of our spectra of SN 1993J from the Lick and Keck
Observatories\footnote{Spectropolarimetry obtained at Lick on days 24,
34, and 45 is discussed by Tran et al. (1997), and is not duplicated
here.} (\S 3), along with a general description of the spectra through
the various phases of evolution for the SN in \S 4.  Detailed analysis
of our spectra for individual phases is discussed elsewhere
(Filippenko, Matheson, \& Ho 1993, hereinafter FMH93; Filippenko,
Matheson, \& Barth 1994, hereinafter FMB94; Matheson et al. 2000,
hereinafter Paper II).

\section{Previous Studies of Supernova 1993J}

The evolution of the light curve of SN 1993J did not follow either of
the two typical paths for SNe II.  SN 1993J did not remain at a
relatively constant brightness after a slight decline from maximum, as
a normal Type II plateau SN would, nor did the brightness decline in
the pattern of a Type II linear SN (for examples of these Type II
light curves, see, e.g., Doggett \& Branch 1985).  Instead, SN 1993J
rose quickly, then rapidly declined for $\sim$1 week, only to brighten
a second time over the next two weeks.  This led to another rapid
decrease in brightness for $\sim$3 weeks, followed by an approximately
exponential decline.  For a complete discussion of the photometry of
SN 1993J, see Okyudo et al. (1993), Schmidt et al. (1993), van
Driel et al. (1993), Wheeler et al. (1993), Benson et al. (1994),
Lewis et al. (1994), Richmond et al. (1994, 1996), Barbon et al. (1995),
Doroshenko et al. (1995), and Prabhu et al. (1995).

The unusual initial behavior of the light curve rapidly led many SN
modelers to conclude that SN 1993J was the result of a core-collapse
explosion in a progenitor that had lost a significant fraction of its
hydrogen envelope, leaving only $\sim 0.1-0.5 M_{\sun}$ of hydrogen.
The original envelope could have been lost through winds (H\"oflich,
Langer, \& Duschinger 1993) from a fairly massive star ($25-30
M_{\sun}$).  Another possibility explored by Hashimoto, Iwamoto, \&
Nomoto (1993; see also Nomoto et al. 1993) is that SN 1993J was the
result of the explosion of an asymptotic giant branch star having
main-sequence mass $M_{ms} \approx 7-10 M_{\sun}$, with a helium-rich
envelope.

A more likely solution is that the progenitor of SN 1993J was a member
of a binary system and the companion had stripped away a considerable
amount of hydrogen.  The progenitor was observed during prior studies
of M81.  Aldering, Humphreys, \& Richmond (1994) analyzed several sets
of pre-existing images and deduced that the photometry was
inconsistent with a single star at the position of SN 1993J.  They
found that the best fit for the progenitor itself was a K0~I star with
$M_{\rm{bol}} \approx -7.8$ mag and $\vr \approx 0.7$ mag.  Cohen,
Darling, \& Porter (1995) derived a similar color from a five-month
series of images of M81 from 1984; there was no apparent variability.

Using the scenario of a star that had been stripped of most of its
hydrogen envelope, Nomoto et al. (1993) and Shigeyama et al. (1994)
found a best fit to the light curve from their model of a $4 M_{\sun}$
helium core, although a range of $3-6 M_{\sun}$ for the core is
reasonable.  The main-sequence mass of the star would have been 15
$M_{\sun}$, while the residual hydrogen envelope is less than $\sim
0.9 M_{\sun}$.  Starting with a star of initial mass of $13-16
M_{\sun}$, Woosley et al. (1994) could reproduce the light curve from
the explosion of a remaining helium core with mass $4.0 \pm 0.5
M_{\sun}$ and hydrogen envelope with mass $0.20 \pm 0.05 M_{\sun}$.  A
similar model by Podsiadlowski et al. (1993) had $0.2 M_{\sun}$ of
hydrogen remaining on a star of initial mass $M_i \approx 15
M_{\sun}$.  Ray, Singh, \& Sutaria (1993) also invoked a binary system
for SN 1993J with a residual hydrogen envelope mass of $0.2 M_{\sun}$.
Utrobin (1994) used an envelope mass of $0.1 M_{\sun}$ remaining on a
$3 M_{\sun}$ helium core from an initial mass of $12 M_{\sun}$.
Bartunov et al. (1994) achieved a good fit to the light curve with a
helium core mass of 3.5 $M_{\sun}$, but a larger hydrogen envelope
($M_{env} \approx 0.9 M_{\sun}$).  Later studies continued to conclude
that a low-mass envelope of hydrogen on a helium core was the most
likely scenario for the progenitor (Young, Baron, \& Branch 1995;
Utrobin 1996).  Intercomparison of two methods also indicated that the
results were robust (Blinnikov et al. 1998).  Houck \& Fransson (1996)
used a non-local thermodynamic equilibrium (NLTE) synthetic spectrum
code to fit nebular spectra and found that the Nomoto et al. (1993)
models could explain the late-time spectra.  They found a best fit
with a 3.2 $M_{\sun}$ helium core with a $0.2-0.4 M_{\sun}$ hydrogen
envelope.  Patat, Chugai, \& Mazzali (1995) also used the late-time
spectra, specifically the H$\alpha$ line, to derive an ionized
hydrogen mass of $0.05-0.2 M_{\sun}$; this is a lower limit to the
envelope mass.

Woosley et al. (1987) had already considered such a possibility for
core-collapse SNe, giving them a new name: SNe IIb.  The low-mass
outer layer of hydrogen would give the initial appearance of a SN II,
but the spectrum would slowly change to one more similar to that of a
SN Ib, dominated by helium lines with the hydrogen either appearing
weakly or completely gone.  Indeed, Nomoto et al. (1993) predicted
that the spectrum of SN 1993J would show this behavior.  This was
first confirmed by Filippenko \& Matheson (1993), followed rapidly by
Schmidt et al. (1993) and Swartz et al. (1993).  A complete analysis
of our optical spectra covering the transformation of SN 1993J from a
Type II to a Type IIb is presented by FMH93.  Other studies of the
early optical spectra include those of Wheeler et al. (1993),
Taniguchi et al. (1993), Garnavich \& Ann (1994), Ohta et al. (1994),
Prabhu et al. (1995), and Metlova et al. (1995).

Jeffery et al. (1994) present an early ultraviolet (UV) spectrum of SN
1993J taken with the \emph{Hubble Space Telescope (HST)} on 1993 April
15.  The other core-collapse SNe that had been observed in the UV to
that point were compared with SN 1993J, and there were striking
differences.  SN 1993J had a relatively smooth UV spectrum and was
more similar to SN 1979C and SN 1980K, both of which are radio sources
and thus likely to have thick circumstellar envelopes (e.g., Weiler et
al. 1986).  The UV spectra of SN 1987A, in contrast, showed broad
absorption features.  The illumination from circumstellar interaction
may reduce the relative strengths of line features compared to the
continuum and thus produce the featureless UV spectra of SNe 1979C,
1980K, and 1993J (Branch et al. 2000).

SN 1993J then evolved fairly rapidly into the nebular phase; our
analysis of this transition is covered by FMB94.  The nebular-phase
spectra were similar to those of a typical SN Ib, but the hydrogen
lines never faded completely.  In fact, H$\alpha$ began to dominate
the spectrum at late times, most likely the result of circumstellar
interaction.  There were several other papers that considered the
nebular-phase spectra (and some relatively late-time spectra).  Lewis
et al. (1994) present the complete La Palma archive covering days 2
through 125.  Li et al. (1994) discuss the nebular-phase spectra
observed from the Beijing Astronomical Observatory.  Barbon et
al. (1995) show the first year of observations from Asiago; the
transformation of the SN from Type II to IIb is evident, as is the
return of H$\alpha$ at late times (by $\sim$ 200 days).  A longer
baseline ($\sim 500$ days) for the spectra is found in the work of
Finn et al. (1995); all the features described above are evident in
their spectra, but the very late-time observations show even more
clearly the presence of circumstellar interaction.

There were optical spectropolarimetric observations of SN 1993J.
Trammell, Hines, \& Wheeler (1993) found a continuum polarization of
$P = 1.6\% \pm 0.1\%$ on day 24 (assuming 1993 March 27.5 as the
explosion date; see below).  Trammell et al. (1993), as well as later
considerations of the same data (H\"oflich 1995; H\"oflich et
al. 1996), argued that this polarization implied an overall asymmetry,
but the source of this asymmetry was undetermined.  The presence of SN
1993J in a binary system was implicated as a potential source for the
asymmetry.  With more epochs of observation (days 7, 8, 11 [$V$-band
only], 24, 30, 31, 32, 34, 45, and 48), Tran et al. (1997) also found
a polarization in the continuum of $\sim$ 1\%, but a different level
for the interstellar polarization.  Nevertheless, they also concluded
that SN 1993J was asymmetric.
It is interesting to note that a subsequent SN IIb, SN 1996cb, showed
substantially similar polarization of its spectra (Wang et al. 2000).

The analysis of individual aspects of the spectra has yielded some
interesting results.  Both Wang \& Hu (1994) and Spyromilio (1994)
found evidence for clumpy ejecta with blueshifted emission lines.
Houck \& Fransson (1996) argue that the lines are not actually
blueshifted, but that contamination from other lines appears to shift
them.  Nonetheless, the lines do show substructure that indicates
clumpy ejecta.  The discussion of the clumpy nature of the ejecta
based on analysis of our spectra is presented in Paper II.

Models of the early spectra could reproduce their overall shape, but
the line strengths were problematic.  Baron et al. (1993) found a
photospheric temperature of $\sim$ 8000 K for day 10, but the
predicted hydrogen and helium lines were too weak, possibly indicating
unusual abundances or non-thermal effects.  A later analysis including the
\emph{HST} UV spectrum was fit well by including enhanced helium
abundance and NLTE effects (Baron, Hauschildt, \& Branch 1994).
Jeffery et al. (1994) also had difficulties fitting line strengths for
transitions that are susceptible to NLTE effects.  Clocchiatti et
al. (1995) studied the early spectra to follow the evolution of color
temperature and to calculate a distance to M81 ($\sim$ 3.5 Mpc) using
the expanding photosphere method (e.g., Eastman, Schmidt, \& Kirshner
1996).  The NLTE treatment of calcium is explored by Zhang \& Wang
(1996), who found a best fit with a reduced calcium abundance.

SN 1993J was also observed quite thoroughly at other wavelengths (see
Wheeler \& Filippenko 1996 for a summary of early results).  Pooley \&
Green (1993) presented a light curve at 2 cm obtained with the Ryle
Telescope.  These data were merged by Van Dyk et al. (1994) with the
VLA data, which included not only observations at 2 cm, but also at
1.3 cm, 3.6 cm, 6 cm, and 20 cm.  Interpreting the data within the
context of standard theories of radio emission from SNe (e.g.,
Chevalier 1982, 1984; Weiler et al. 1986; Weiler, Panagia, \& Sramek
1990), Van Dyk et al. (1994) concluded that the radio absorbing
material around the SN is inhomogeneous and that the circumstellar
density profile is flatter than that expected for a constant mass-loss
rate, constant-velocity wind.  The flatter profile could be the result
of decreasing mass-loss rate or increasing wind velocity.  SN 1993J
was also observed using very long baseline interferometry (VLBI).
Early studies indicated either a disk-like structure or an optically
thick shell (Marcaide et al. 1994), and other observations found SN
1993J to be circularly symmetric and expanding linearly (Bartel et
al. 1994).  Later measurements found a spherically symmetric shell
(Marcaide et al. 1995).  After years of observation using VLBI,
Marcaide et al. (1997) still observed a symmetric shell, but the SN
expansion had apparently begun to decelerate.  With an even longer
baseline, Bartel et al. (2000) also found that the expansion had begun
to decelerate, as well as revealing some structure in the ejecta.  As
these studies probe different regions than the polarimetric
observations described above, the spherical symmetry revealed by the
VLBI images is not inconsistent with the asymmetry deduced from
polarimetry.

Soft X-rays were detected from SN 1993J six days after the explosion
by \emph{ROSAT} (Zimmerman et al. 1994) and two days later by
\emph{ASCA} (Kohmura et al. 1994).  Both observatories followed the SN
for several months, recording a gradual decrease in X-ray flux as well
as a softening of the energy spectrum.  The \emph{Compton Gamma-Ray
Observatory} detected SN 1993J with OSSE, and hard X-rays ($>$ 50 keV)
were observed on at least two epochs (Leising et al. 1994).  All three
of these groups, as well as Suzuki \& Nomoto (1995) and Fransson,
Lundqvist, \& Chevalier (1996), interpreted these observations as
indications of circumstellar interaction, with the X-rays coming from
either the shocked wind material or the reverse-shocked SN ejecta.
Both Patat et al. (1995) and Houck \& Fransson (1996) concluded that
the late-time optical spectra could only be powered by a circumstellar
interaction; radioactive decay alone was not enough.

Although SN 1993J has provided the best observational evidence for the
transformation of a SN from one type to another, there have been other
examples.  The early spectra of SN 1987K showed hydrogen lines, but
the late-time spectra more closely resembled those of SNe Ib
(Filippenko 1988).  The transition itself was not observed, occurring
while SN 1987K was in conjunction with the Sun.  SN 1996cb underwent a
very similar metamorphosis from SN II to SN Ib; Qiu et al. (1999)
present a complete spectroscopic record of the transformation.  In
addition, there were some suggestions of hydrogen in spectra of the
Type Ic SN 1987M (Jeffery et al. 1991; Filippenko 1992) and the SN Ic
1991A (and perhaps SN Ic 1990aa; Filippenko 1992).  SN 1993J is
clearly a significant object in the study of SNe.  By providing a link
between SNe II and SNe Ib, it has strengthened the argument that SNe
Ib (and, by extension, SNe Ic) are also core-collapse events.

\section{Observations and Reductions}

\subsection{Low-Dispersion Spectra}

Low-dispersion spectra of SN 1993J were obtained with the Kast double
spectrograph (Miller \& Stone 1993) at the Cassegrain focus of the
Shane 3-m reflector at Lick Observatory and with the Low Resolution
Imaging Spectrometer (LRIS; Oke et al. 1995) at the Cassegrain focus
of the Keck 10-m telescopes (both Keck~I and Keck~II were used).  The
Kast spectrograph has Reticon $400 \times 1200$ pixel CCDs in both
cameras, while LRIS has a single Tektronix $2048 \times 2048$ pixel
CCD.  The spatial scale for the Kast CCD was 0.$\arcsec$8 per pixel;
the LRIS CCD was binned in the spatial direction, yielding
0.$\arcsec$43 per pixel.  At Lick, the slit width was generally
2$\arcsec$, but 8$\arcsec$ observations were also taken on potentially
photometric nights to provide an absolute flux scale; the 2$\arcsec$
data were scaled to the flux level of the 8$\arcsec$ observation.  The
8$\arcsec$ observations were done only when SN 1993J was reasonably
bright (until about day 200).  The LRIS slit width was typically
1$\arcsec$, but 0.$\arcsec$7 widths were also used when conditions
allowed or when attempting to obtain better spectral resolution.
Various gratings and grisms were utilized, yielding resolutions (full
width at half maximum, FWHM) ranging from 2.5 \AA\ to 15 \AA.  Details
of the exposures are given in Table 1.  Most of the spectra were taken
with the slit oriented at, or near (within 10$^{\circ}$), the
parallactic angle (Filippenko 1982), but exceptions are noted in Table
1.  We follow Lewis et al. (1994) in adopting 1993 March 27.5 (JD
2,449,074) as the date of explosion.

Standard CCD processing and spectrum extraction were accomplished with
VISTA (Terndrup, Lauer, \& Stover 1984) through day 553 and
IRAF\footnote{IRAF is distributed by the National Optical Astronomy
Observatories, which are operated by the Association of Universities
for Research in Astronomy, Inc., under cooperative agreement with the
National Science Foundation.} for day 670 and thereafter (day 56 was
also processed using IRAF).  Optimal extraction was used for the IRAF
reductions (Horne 1986).  The wavelength scale was established using
low-order polynomial fits to calibration lamps of
{\small\rmfamily{He-Hg-Cd-Ne-Ar}} (Lick) or
{\small\rmfamily{He-Ne-Kr-Ar}} (Keck).  The typical root-mean-square
(rms) deviation for the wavelength solution was $0.1-0.5$ \AA,
depending on the resolution for the particular exposure.  Final
adjustments to the wavelength scale were obtained by using the
background sky regions to provide an absolute scale.  We employed our
own routines to flux calibrate the data; comparison stars are listed
in Table 1.  Particular care was taken to remove telluric absorption
features through division by an intrinsically featureless spectrum,
where possible (Wade \& Horne 1988; see also Paper II).  The flux
standard was routinely employed for this purpose.

As most of the spectra were observed with the position angle oriented
along or near the parallactic angle, the relative spectrophotometry is
quite good.  For the nights during which an 8$\arcsec$ slit width
exposure was taken to provide an absolute flux scale, we checked our
fluxes against the $BVRI$ photometry of Richmond et al. (1996).  Days
16, 45, and 209 were not photometric.  Days 17, 18, 19, 34, 56, and 93
all agreed quite well with the photometry (within $\sim$ 5\% in $B$
and $V$, slightly larger deviations in $R$ and $I$, probably due to
the differences between the observed passband and the assumed passband
used to calculate fluxes from the spectra).  The flux of the day 109
spectrum has large deviations from the broad-band photometry ($\sim$
20\%).  That night was our first attempt to observe SN 1993J ``under
the pole'' with the equatorially mounted Shane 3-m telescope, and we
attribute the fluxing errors to the difficulties arising from this
complication to the observing program.  Subsequent uses of the
telescope ``under the pole'' were more successful, and days 123 and
139 are in fairly good agreement with the photometry, although they do
differ in $B$.  This may be the result of difficulties aligning the
slit along the parallactic angle for these observations.  Day 167
agrees with the photometry to within $\sim$ 10\%, but day 182 shows
much larger differences.  The spectrum had become almost completely
nebular by this stage, so the effects of strong emission lines falling
near the edges of passbands may explain the discrepancies.  For all of
the other spectra, we had no absolute calibrators.

\subsection{High-Dispersion Spectrum}

A single high-dispersion spectrum was obtained on 1994 April 14 UT
(day 383) with the HIRES echelle spectrometer (Vogt 1992, 1994) on the
Keck~I 10-m telescope.  The HIRES detector is a Tektronix $2048 \times
2048$ pixel CCD.  The setup for these observations encompassed the
range $4240-6720$ \AA\ in 31 spectral orders.  Beyond $\sim$~5100 \AA,
small gaps in the wavelength coverage appear because the CCD was too
small to span the progressively wider orders.  A KV408 order-blocking
filter was used to eliminate second-order blue light.  The ``C5'' slit
decker (1.$\arcsec$15 $\times$ 7$\arcsec$) was utilized to prevent
overlapping orders and to ensure adequate sky background, yielding a
spectral resolution of $R = 38,000$.  The HIRES chip was binned in the
spatial direction, with 0.$\arcsec$41 per pixel.  There were three
exposures of SN 1993J, two with integration times of 2700 s and one
with an integration time of 1800 s.  The seeing was $\sim$
1.$\arcsec$2 and the night was photometric.  At the time, HIRES had
neither an image rotator nor an atmospheric dispersion compensator, so
differential light losses may affect the spectrum; the position angle
of the slit was not at the parallactic angle (Filippenko 1982).  For
all three observations, SN 1993J was at an airmass of $\sim$ 1.5, so
the effects are small, especially redward of 4500 \AA.

Once again, IRAF was used for standard CCD processing and spectrum
extraction.  We extracted one-dimensional spectra by summing the five
pixels centered on a polynomial fit to the centroid of the light
distribution along the dispersion, yielding an effective aperture of
1.$\arcsec$15 $\times$ 2.$\arcsec$05.  A third-order fit to the
thorium-argon comparison lamp spectra provided a wavelength solution
with a dispersion of $\sim$ 0.003 \AA.  No attempt was made to flux
calibrate the data.  The sdG star HD 84937 (Oke \& Gunn 1983) was
observed for use in the identification and removal of telluric lines.
In order to preserve the large-scale shape of the lines, several
different attempts were made using the observed spectrum of HD 84937
and featureless regions of the spectrum of SN 1994I (see Ho \&
Filippenko [1995] for a discussion of the SN 1994I spectrum observed
on this night).  The corrections to the continuum shape made with SN
1994I provided the best representation when compared with
low-dispersion spectra taken on approximately the same date.  Residual
errors remain in the overall shape of individual orders, but the
small-scale structure is reproduced accurately.

\section{Results and Discussion}

\subsection{Days 3 to 34}

Spectra from days 3 to 34 are shown in Figure 1.  The first spectrum
obtained at Lick Observatory was taken on 1993 March 30, day 3.  It
shows a blue, nearly featureless continuum.  There are some broad
undulations that may be incipient P-Cygni features of H$\alpha$ and
\ion{He}{1} $\lambda$5876, but the presence of reduction artifacts
makes interpretation problematic.  Clocchiatti et al. (1995) found a
best-fit blackbody curve for this spectrum that indicated a
temperature of $\sim$ 30,000 K, along with $A_V \approx 0.7$ mag.  The
day 3 spectrum also contained narrow (unresolved) emission features of
\ion{He}{2} $\lambda$4686, [\ion{Fe}{10}] $\lambda$6374, and
H$\alpha$.  Benetti et al. (1994) also observed [\ion{Fe}{14}]
$\lambda$5303.  In addition, there are narrow absorption components
of \ion{Ca}{2}~H\&K and \ion{Na}{1}~D.  The observed wavelengths of
the narrow emission features allowed us to derive a relative
heliocentric velocity for the SN of $-140$ km s$^{-1}$, well in
agreement with other results (e.g., Vladilo et al. 1993, $-135$ km
s$^{-1}$).  All spectra presented herein have had this velocity
removed.  More detailed studies using high-resolution spectrographs
that discuss the narrow lines in the spectra of SN 1993J include those
of Benetti et al. (1994), Vladilo et al. (1993, 1994), and Bowen et
al. (1995).

As the SN cools, line structure begins to appear in the spectra.  By
days 16 through 19, SN 1993J resembles a relatively typical SN II,
although some details of the lines are slightly unusual.  The emission
component of the H$\alpha$ P-Cygni line has a flat top.  Figure 2
shows the day 19 spectrum on a linear flux-density scale with some
line identifications.  Overall, this spectrum is very much like that
of most SNe II.  The spectra from days 32 through 34 start to show
that SN 1993J is not a typical SN II.  A broad notch develops in the
H$\alpha$ profile, and the emission component begins to split in two.
In addition, the profile associated with \ion{He}{1} $\lambda$5876 and
\ion{Na}{1}~D strengthens, especially in the absorption component.  As
described above, this is interpreted as the onset of a phase wherein
the \ion{He}{1} lines grow in prominence, with the notch in H$\alpha$
being the P-Cygni profile of \ion{He}{1} $\lambda$6678.  Note that the
narrow feature seen in the H$\alpha$ emission profile of earlier
spectra at the wavelength of the helium notch is most likely telluric
absorption (see Paper II for details).  Another indication of helium
is the P-Cygni profile of \ion{He}{1} $\lambda$7065 that also appears
on day 32 in our spectra.  Other \ion{He}{1} lines are blended with
other lines or are weak, and so are less convincing than $\lambda$6678
and $\lambda$7065.

The notch in H$\alpha$ is not obvious in the day 22 spectra of Barbon
et al. (1995) or Lewis et al. (1994), or in the day 23 spectrum of
Prabhu et al. (1995).  It does appear in the day 25 spectra of Prabhu
et al. (1995) and Barbon et al. (1995); they also have a day
26 spectrum with the notch.  The notch is clearly present in the day
27 spectrum of Finn et al. (1995).  (Note that the day 26 spectrum of
Lewis et al. [1994] appears to have some reduction and/or observation
errors; it is very different from the spectra of the other groups at
similar times, including their own from day 22.)

\subsection{Days 45 to 109}

The helium features continue to strengthen and the metamorphosis of
SN 1993J from a Type II to IIb becomes readily apparent in days 45
through 109, shown in Figure 3.  The day 45 spectrum (Figure 4)
exhibits almost the entire \ion{He}{1} series of lines.  The most
obvious lines remain \ion{He}{1} $\lambda$5876, $\lambda$6678, and
$\lambda$7065, although $\lambda$5876 is probably contaminated by
\ion{Na}{1}~D.  Other \ion{He}{1} lines are also blended with other
features, including $\lambda$7281 (contaminated by [\ion{Ca}{2}]
$\lambda\lambda$7291, 7324), $\lambda$5015 (blended with \ion{Fe}{2}
$\lambda$5018), $\lambda$4921 (affected by H$\beta$ and \ion{Fe}{2}
$\lambda$4924), and $\lambda$4471 (blended with H$\gamma$).

By day 89, the helium lines begin to weaken in comparison with the
rest of the spectrum, and they are effectively gone by day 109,
although a strong absorption due to \ion{He}{1} $\lambda$5876 and
\ion{Na}{1}~D remains.  This is similar to the late-time behavior of a
typical SN Ib (e.g., Gaskell et al. 1986; Branch, Nomoto, \& Filippenko
1991).  Although not quite at the fully nebular phase, some
nebular lines begin to appear at these times, including \ion{Mg}{1}]
$\lambda$4571, [\ion{O}{1}] $\lambda\lambda$6300, 6364, and
[\ion{Ca}{2}] $\lambda\lambda$7291, 7324.

\subsection{Days 123 to 266}

The spectra from days 123 through 226 (Figure 5) show fairly little
evolution.  The nebular lines strengthen relative to the continuum
over time, resulting in an almost purely emission-line spectrum.  The
most dramatic aspect of these spectra is the continuing weakness of
H$\alpha$.  Qualitatively following the relative strength of
[\ion{O}{1}] $\lambda\lambda$6300, 6364 and H$\alpha$ over these days
shows the oxygen lines evolving to dominate the hydrogen strength.
(Blending of the lines makes a quantitative ratio of these features
very difficult.)  The spectrum from day 182 is shown in Figure 6
with some line identifications.  Aside from the weak H$\alpha$, this
could easily be the nebular spectrum of a SN Ib or SN Ic.

\subsection{Days 286 to 553}

The nebular lines begin to fade away during days 286 through 553
(Figure 7).  The most prominent lines in the spectra on day 286 ---
\ion{Mg}{1}] $\lambda$4571, [\ion{O}{1}] $\lambda\lambda$6300, 6364,
[\ion{Ca}{2}] $\lambda\lambda$7291, 7324, and the \ion{Ca}{2} near-IR
triplet --- are virtually gone by days 473 and 523.  The H$\alpha$
line, though, remains at a fairly constant strength as the
[\ion{O}{1}] doublet decreases (see Figure 4 of FMB94).  H$\alpha$ is
comparable to [\ion{O}{1}] $\lambda\lambda$6300 6364 in strength by
days 387 and 433, and the oxygen line is reduced to a small feature on
the blue shoulder of the H$\alpha$ line by day 553.

The day 387 spectrum is shown in Figure 8.  Note the development of
the box-like H$\alpha$ profile in comparison with day 182 (Figure 6).
The presence of the box-like profile for H$\alpha$ on day 553, as well
as for other lines (\ion{He}{1} $\lambda$5876 and \ion{Na}{1}~D,
possibly [\ion{O}{3}] $\lambda$4363 and [\ion{O}{3}]
$\lambda\lambda$4959, 5007), indicates that a new emission phase has
begun for SN 1993J---one dominated by the effects of circumstellar
interaction.  This profile for H$\alpha$ is discernible as early as
day 298, perhaps even day 226, but it does not dominate the emission
until day 473 and beyond.  These emission profiles imply a roughly
spherical distribution for the source material, probably a
geometrically thin shell, similar to that discussed above in the
context of late-time radio observations.  The nature of the spectra is
very similar to that described by Chevalier \& Fransson (1994).
Another interesting aspect of the day 387 spectrum (and earlier
spectra) is the presence of small-scale features in some of the
emission lines.  They are even more evident in Figure 9, wherein the
region of the spectrum containing [\ion{O}{1}] $\lambda\lambda$6300
6364 and H$\alpha$ is shown in greater detail, along with the HIRES
spectrum obtained four days earlier.  A full exploration of the clumps
exhibited in the spectra of SN 1993J is presented in Paper II.

\subsection{Days 670 to 2454}

Circumstellar interaction continues to dominate the spectra of SN
1993J up to the most recent observations.  Spectra from days 670
through 2454 are shown in Figure 10.  The box-like profiles are
especially obvious in the day 976 spectrum (Figure 11).  While these
profiles are present earlier, the high S/N ratio of this spectrum
allows one to see clearly the ``double-box'' created by the
overlapping lines of \ion{He}{1} $\lambda$5876 and \ion{Na}{1}~D.  

One striking change that occurs during this period is highlighted by
the spectra from day 976 (Figure 11) and day 1766 (Figure 12).  There
appear to be narrower emission features on top of the boxy profiles
for [\ion{O}{3}] $\lambda\lambda$4959, 5007 and [\ion{O}{2}]
$\lambda\lambda$7319, 7330 on day 1766 that are not obvious in the day
976 spectrum.  Unfortunately, there is a large gap in the coverage at
this point, so it is not known when these features appeared.  They do
show up at the same relative velocity in each profile (including the
[\ion{O}{1}] $\lambda\lambda$6300, 6364 doublet), indicating that they
are related to the underlying oxygen lines.  We interpret these
features as the blue and red peaks of a double-horned profile due to a
somewhat flattened, perhaps even disk-like emission source, with the
red peak attenuated by absorption (see Paper II for details).  They
are superposed on the box-like profile from a roughly spherically
distributed source.  This is especially intriguing in light of the
polarization measurements done at early times (Trammell et al. 1993;
Tran et al. 1997).

With the knowledge of the day 1766 spectrum, the incipient beginnings
of these features are discernible in the day 976 spectrum.  In fact,
the day 976 spectrum shows the weak, narrow remnant of the once
prominent [\ion{O}{1}] $\lambda$6300 line still present, but fading
in comparison with the blue component of the two-horned profile of
[\ion{O}{1}].  These features persist, being strongly evident in the
day 2454 spectrum (Figure 13).  

The only other significant change in the later spectra is the
strengthening of [\ion{O}{3}] $\lambda\lambda$4959, 5007 relative to
[\ion{O}{3}] $\lambda$4363, indicating a drop in the density of the
oxygen-emitting regions.  The [\ion{O}{1}] $\lambda\lambda$6300, 6364
doublet begins to grow again relative to H$\alpha$.  These changes are
especially clear when comparing the day 2454 spectrum (Figure 13) with
the day 976 spectrum (Figure 11).  Note that there is very little
change between days 2176 and 2454 (cf. Figure 10).  The overall spectra
exhibit almost solely the emission from circumstellar interaction.
The line-intensity ratios are generally similar to the predictions of
Chevalier \& Fransson (1994), although some differ.  Details of this
analysis of the late-time spectral lines is presented in Paper II.

\section{Conclusions}

We have presented the complete existing collection of low-dispersion
spectra of SN 1993J obtained at Lick and Keck Observatories, as well
as one high-dispersion spectrum from Keck.  These 42 low-dispersion
spectra, representing coverage from day 3 after explosion to day 2454,
document thoroughly the distinctive characteristics of SN 1993J.  The
early spectra show a slightly unusual Type II event, followed by the
appearance of strong helium absorption lines.  After SN 1993J
underwent this metamorphosis from Type II to IIb, the nebular phase
developed rapidly, appearing as a typical late-time SN Ib or SN Ic,
but with a weak residual H$\alpha$ line.  As the nebular lines
weakened, new line profiles emerged with box-like shapes implying
circumstellar interaction with a spherical shell-like distribution.
At even later times, double-horned profiles appeared on top of the box
profiles, indicating the presence of a somewhat flattened or disk-like
morphology along with the shell.  As SN 1993J appears to be fading
slowly (see Paper II), we plan to continue monitoring it
spectroscopically for years to come.  The metamorphosis in our spectra
was discussed by FMH93, while the nebular spectra were analyzed by
FMB94.  Our study of the detailed line structure at relatively early
times and the circumstellar interaction phase is presented in Paper
II.

\acknowledgments

This research was supported by NSF grants AST-9115174 and AST-9417213
to A.V.F.  We are grateful to the staffs of the Lick and Keck
Observatories for help with the observations; we are especially
thankful to the Lick staff for their heroic efforts in reconfiguring
the hardware and software of the equatorial Shane 3-m telescope to
allow us to observe ``under the pole'' during the summer months of
1993 and 1994.  William and Marina Kast provided a generous gift that
led to the construction of the double spectrograph on the 3-m
telescope that was used for most of the observations reported here.
The W. M. Keck Observatory is operated as a scientific partnership
among the California Institute of Technology, the University of
California, and NASA. The Observatory was made possible by the
generous financial support of the W. M. Keck Foundation.  Robert
H. Becker and Richard L. White graciously provided the 1993 June 24
(day 89) spectrum.  We thank Alison Coil, Ryan Chornock, Isobel Hook,
Chien Peng, Saul Perlmutter, Adam Riess, Hien Tran, and Schuyler Van
Dyk for assistance with some of the observations and reductions.


\clearpage



\figcaption{Spectra of SN 1993J from days 3 to 34, assuming explosion
on 1993 March 27.5 (JD 2,449,074).  The flux units are $-2.5$ log
$f_{\lambda} - 21.10$, following the definition of Space Telescope
(ST) magnitudes (e.g., Koorneef et al. 1986).  ST magnitudes are
analogous to AB magnitudes ($-2.5$ log $f_{\nu} - 48.60$; Oke \& Gunn
1983), with the zero point yielding monochromatic magnitudes for Vega
in the Johnson $V$ passband of $\sim$ 0.  The following constants have
been added to the individual spectra (from top to bottom): +1.7,
$-0.5$, 0.0, 1.0, 2.0, 2.3, 3.5, and 5.3.  Telluric absorption features
are indicated for the day 3 spectrum from which they could not be
properly removed.  In this, and all subsequent figures, the systemic
heliocentric velocity of $-140$ km s$^{-1}$ has been removed.\label{fig1}}

\figcaption{Spectrum of SN 1993J from day 19 (1993 April 15) with line
identifications.  The flux-density scale is linear and has been scaled
to match the $V$-band photometry of Richmond et al. (1996).  The
identification of the line at $\sim$ 4430 \AA\ is uncertain, but it may
be \protect\ion{Ba}{2} $\lambda$4554 (see, e.g., Williams 1987;
Turatto et al. 1998; Fassia et al. 1998).\label{fig2}}

\figcaption{Spectra of SN 1993J from days 45 to 109, with date of
explosion and flux units as in Figure 1.  The following constants have
been added to the individual spectra (from top to bottom): 0.0, 1.5,
3.0, 4.5, 6.5, 8.3, 11.3, and 13.5.  Telluric absorption features are
indicated for the day 91 spectrum from which they could not be
properly removed.
\label{fig3}}

\figcaption{Spectrum of SN 1993J from day 45 (1993 May 11) with line
identifications.  The flux-density scale is linear and has been scaled
to match the $V$-band photometry of Richmond et al. (1996).\label{fig4}}

\figcaption{Spectra of SN 1993J from days 123 to 226, with date of
explosion and flux units as in Figure 1.  The following constants have
been added to the individual spectra (from top to bottom): 0.0, 2.0,
4.0, 6.5, 9.0, 11.0, 15.0, and 19.0.  Noisy spectra have been clipped for
clarity.
\label{fig5}}

\figcaption{Spectrum of SN 1993J from day 182 (1993 September 25) with line
identifications.  The flux-density scale is linear and has been scaled
to match the $V$-band photometry of Richmond et al. (1996).\label{fig6}}

\figcaption{Spectra of SN 1993J from days 286 to 553, with date of
explosion and flux units as in Figure 1.  The following constants have
been added to the individual spectra (from top to bottom): 0.0, 2.5,
5.5, 7.5, 12.5, 15.5, 19.0, 22.0, and 25.0.  Noisy spectra have been
clipped for clarity.
\label{fig7}}

\figcaption{Spectrum of SN 1993J from day 387 (1994 April 18) with line
identifications.  The flux-density scale is linear and has been scaled
to match the $V$-band photometry of Richmond et al. (1996).\label{fig8}}

\figcaption{Low-resolution spectrum of SN 1993J on day 387 compared
with a high-resolution spectrum on day 383 (1994 April 14).  The
high-resolution spectrum has been binned to 0.25 \AA/pixel for
clarity.  The global shape of each order of the high-resolution
spectrum is not necessarily accurate.  The substructure seen in the
low-dispersion spectrum is reflected in the high-dispersion spectrum,
and there is no apparent structure hidden by the lower resolution of
the low-dispersion spectrum.\label{fig9}}

\figcaption{Spectra of SN 1993J from days 670 to 2454, with date of
explosion and flux units as in Figure 1.  The following constants have
been added to the individual spectra (from top to bottom): 0.0, 3.5,
6.5, 7.5, 10.5, 15.0, 17.5, 20.0, and 22.5.  Noisy spectra have been
clipped for clarity.
\label{fig10}}

\figcaption{Spectrum of SN 1993J from day 976 (1995 November 28) with line
identifications.  The flux-density scale is linear.\label{fig11}}

\figcaption{Spectrum of SN 1993J from day 1766 (1998 January 26) with line
identifications.  The flux-density scale is linear.\label{fig12}}

\figcaption{Spectrum of SN 1993J from day 2454 (1999 December 15) with line
identifications.  The flux-density scale is linear.  There is some
second-order contamination of the spectrum beyond $\sim$~7600 \AA,
although it is not significant; as Figure 10 shows, there is little
structure in this region of the spectrum on earlier days.\label{fig13}}

\clearpage



\begin{thebibliography}{}
\bibitem{} Aldering, G., Humphreys, R. M., \& Richmond, M. 1994, \aj,
107, 662
\bibitem{} Barbon, R., Benetti, S., Cappellaro, E., Patat, F.,
Turatto, M., \& Iijima, T. 1995, \aaps, 110, 513
\bibitem{} Barbon, R., Ciatti, F., \& Rosino, L. 1973, \aap, 25, 241
\bibitem{} Baron, E., Hauschildt, P. H., \& Branch, D. 1994, \apj,
426, 334
\bibitem{} Baron, E., Hauschildt, P. H., Branch, D., Wagner, R. M.,
Austin, S. J., Filippenko, A. V., \& Matheson, T. 1993, \apj, 416, L21
\bibitem{} Bartel, N., et al. 2000, Science, 287, 112
\bibitem{} Bartel, N., et al. 1994, \nat, 368, 610
\bibitem{} Bartunov, O. S., Blinnikov, S. I., Pavlyuk, N. N., \&
Tsvetkov, D. Yu. 1994, \aap, 281, L53
\bibitem{} Benetti, S., Patat, F., Turatto, M., Contarini, G.,
Gratton, R., \& Cappellaro, E. 1994, \aap, 285, L13
\bibitem{} Benson, P. J., et al. 1994, \aj, 107, 1453
\bibitem{} Blinnikov, S. I., Eastman, R., Bartunov, O. S., Popolitov,
V. A., \& Woosley, S. E. 1998, \apj, 496, 454
\bibitem{} Branch, D., Jeffery, D. J., Blaylock, M., \& Hatano,
K. 2000, \pasp, 112, 217
\bibitem{} Branch, D., Nomoto, K., \& Filippenko, A. V. 1991,
Comm. Ap., 15, 221
\bibitem{} Bowen, D. V., Roth, K. C., Blades, J. C., \& Meyer,
D. M. 1995, \apj, 420, L71
\bibitem{} Chevalier, R. A. 1982, \apj, 259, 302
\bibitem{} Chevalier, R. A. 1984, \apj, 285, L63
\bibitem{} Chevalier, R. A., \& Fransson, C. 1994, \apj, 420, 268
\bibitem{} Clocchiatti, A., Wheeler, J. C., Barker, E. S., Filippenko,
A. V., Matheson, T., \& Liebert, J. W. 1995, \apj, 446, 167
\bibitem{} Cohen, J. G., Darling, J., \& Porter, A. 1995, \aj, 110, 308
\bibitem{} Doggett, J. B., \& Branch, D. 1985, \aj, 90, 2303
\bibitem{} Doroshenko, V. T., Efimov, Yu. S., \& Shakhovskoi,
N. M. 1995, Ast. Lett., 21, 513
\bibitem{} Eastman, R. G., Schmidt, B. P., \& Kirshner, R. 1996, \apj,
466, 911
\bibitem{} Fassia, A., Meikle, W. P. S., Geballe, T. R., Walton,
N. A., Pollacco, D. L., Rutten, R. G. M., \& Tinney, C. 1998, \mnras,
299, 150
\bibitem{} Filippenko, A. V. 1982, \pasp, 94, 715
\bibitem{} Filippenko, A. V. 1988, \aj, 96, 1941
\bibitem{} Filippenko, A. V. 1992, \apj, 384, L37
\bibitem{} Filippenko, A. V. 1997, \araa, 35, 309
\bibitem{} Filippenko, A. V., \& Matheson, T. 1993, \iaucirc\ 5787
\bibitem{} Filippenko, A. V., Matheson, T., \& Barth, A. J. 1994, \aj,
108, 2220 (FMB94)
\bibitem{} Filippenko, A. V., Matheson, T., \& Ho, L. C. 1993, \apj,
415, L103 (FMH93)
\bibitem{} Filippenko, A. V., Treffers, R. R., Paik, Y., Davis, M., \&
Schlegel, D. 1993, \iaucirc\ 5731
\bibitem{} Finn, R. A., Fesen, R. A., Darling, G. W., Thorstensen,
J. R., \& Worthey, G. S. 1995, \aj, 110, 300
\bibitem{} Fransson, C., Lundqvist, P., \& Chevalier, R. A. 1996,
\apj, 461, 993
\bibitem{} Freedman, W. L., et al. 1994, \apj, 427, 628
\bibitem{} Garnavich, P., \& Ann, H. B. 1993, \iaucirc\ 5731;
corrigendum:  \iaucirc\ 5733
\bibitem{} Garnavich, P. M., \& Ann, H. B. 1994, \aj, 108, 1002
\bibitem{} Gaskell, C. M., Cappellaro, E., Dinerstein, H. L., Garnett,
D. R., Harkness, R., P., \& Wheeler, J. C. 1986, \apj, 306, L77
\bibitem{} Hashimoto, M., Iwamoto, K., \& Nomoto, K. 1993, \apj, 414,
L105
\bibitem{} Ho, L. C., \& Filippenko, A. V. 1995, \apj, 444, 165
[Erratum 1996, 463, 818]
\bibitem{} H\"oflich, P. 1995, \apj, 440, 821
\bibitem{} H\"oflich, P., Langer, N., \& Duschinger, M. 1993, \aap,
275, L29
\bibitem{} H\"oflich, P., Wheeler, J. C., Hines, D. C., \& Trammell,
S. R. 1996, \apj, 459, 307
\bibitem{} Horne, K. 1986, \pasp, 98, 609
\bibitem{} Houck, J. C., \& Fransson, C. 1996, \apj, 456, 811
\bibitem{} Jeffery, D. J., Branch, D., Filippenko, A. V., \& Nomoto, K.
1991, \apj, 377, L89 
\bibitem{} Jeffery, D. J., et al. 1994, \apj, 421, L27
\bibitem{} Kohmura, Y., et al. 1994, \pasj, 46, L157
\bibitem{} Koorneef, J., Bohlin, R., Buser, R., Horne, K., \&
Turnshek, D. 1986, in Highlights of Astronomy, Vol. 7,
ed. J. P. Swings (Dordrecht:  Reidel), p. 833
\bibitem{} Leising, M. D., et al. 1994, \apj, 431, L95
\bibitem{} Lewis, J. R., et al. 1994, \mnras, 266, L27
\bibitem{} Li, A., Hu, J., Wang, L., Jiang, X., \& Li, H. 1994, \apss,
211, 323
\bibitem{} Marcaide, J. M., et al. 1994, \apj, 424, L25
\bibitem{} Marcaide, J. M., et al. 1995, \nat, 373, 44
\bibitem{} Marcaide, J. M., et al. 1997, \apj, 486, L31
\bibitem{} Massey, P., \& Gronwall, C. 1990, \apj, 358, 344
\bibitem{} Matheson, T., Filippenko, A. V., Ho, L. C., Barth, A. J.,
\& Leonard, D. C. 2000, \aj, in press (Paper II)
\bibitem{} Metlova, N. V., Tsvetkov, D. Yu., Shugarov, S. Yu., Esipov,
V. F., \& Pavlyuk, N. N. 1995, Ast. Lett., 21, 598
\bibitem{} Miller, J. S., \& Stone, R. P. S. 1993, Lick
Obs. Tech. Rep., No. 66 
\bibitem{} Nomoto, K., Suzuki, T., Shigeyama, T., Kumagai, S.,
Yamaoka, H., \& Saio, H. 1993, \nat, 364, 507
\bibitem{} Ohta, K., Maemura, H., Ishigaki, T., Aoki, K., \& Ohtani,
H. 1994, \pasj, 46, 117
\bibitem{} Oke, J. B., et al. 1995, \pasp, 107, 375
\bibitem{} Oke, J. B., \& Gunn J. E. 1983, \apj, 266, 713
\bibitem{} Okyudo, M., Kato, T., Ishida, T., Tokimasa, N., \& Yamaoka,
H. 1993, \pasj, 45, L63
\bibitem{} Patat, F., Chugai, H., \& Mazzali, P. A. 1995, \aap, 299,
715
\bibitem{} Podsiadlowski, Ph., Hsu, J. J. L., Joss, P. C., \& Ross,
R. R. 1993, \nat, 364, 509
\bibitem{} Pooley, G. G., \& Green, D. A. 1993, \mnras, 264, L17
\bibitem{} Prabhu, T. P., et al. 1995, \aap, 295, 403
\bibitem{} Qiu, Y., Li, W., Qiao, Q., \& Hu, J., 1999, \aj, 117, 736
\bibitem{} Ray, A., Singh, K. P., \& Sutaria, F. K. 1993,
Astrophys. Astron., 14, 53
\bibitem{} Richmond, M. W., Treffers, R. R., Filippenko, A. V., Paik,
Y., Leibundgut, B., Schulman, E., \& Cox, C. V. 1994, \aj, 107, 1022
\bibitem{} Richmond, M. W., Treffers, R. R., Filippenko, A. V., \&
Paik, Y. 1996, \aj, 112, 732
\bibitem{} Ripero, J., Garcia, F., \& Rodriguez, D. 1993, \iaucirc\ 5731
\bibitem{} Schmidt, B. P., et al. 1993, \nat, 364, 600
\bibitem{} Shigeyama, T., Suzuki, T., Kumagai, S., Nomoto, K., Saio,
H., \& Yamaoka, H. 1994, 420, 341
\bibitem{} Spyromilio, J. 1994, \mnras, 266, L61
\bibitem{} Stone, R. P. S. 1977, \apj, 218, 767
\bibitem{} Suzuki, T., \& Nomoto, K. 1995, \apj, 455, 658
\bibitem{} Swartz, D. A., Clocchiatti, A., Benjamin, R., Lester,
D. F., \& Wheeler, J. C. 1993, \nat, 365, 232
\bibitem{} Taniguchi, Y., Murayama, T., Sato, Y., Yadoumaru, Y.,
Ohyama, Y., Kosugi, G., Yoshida, M., \& Kurakami, T. 1993, \pasj, 45,
L43
\bibitem{} Terndrup, D. M., Lauer, T. R., \& Stover, R. 1984, Lick
Obs. Tech. Rep., No. 33
\bibitem{} Trammell, S. R., Hines, D. C., \& Wheeler, J. C. 1993,
\apj, 414, L21
\bibitem{} Tran, H. D., Filippenko, A. V., Schmidt, G. D., Bjorkman,
K. S., Jannuzi, B. T., \& Smith, P. S. 1997, \pasp, 109, 489
\bibitem{} Turatto, M., et al. 1998, \apj, 498, L129
\bibitem{} Utrobin, V. 1994, \aap, 281, L89
\bibitem{} Utrobin, V. P. 1996, \aap, 306, 219
\bibitem{} van Driel, W., et al. 1993, \pasj, 45, L59
\bibitem{} Van Dyk, S. D., Weiler, K. W., Sramek, R. A., Rupen, M. P.,
\& Panagia, N. 1994, \apj, 324, L115
\bibitem{} Vladilo, G., Centuri\'on, M., de Boer, K. S., King, D. L.,
Lipman, K., Stegert, J. S. W., Unger, S. W., \& Walton, N. A. 1993,
\aap, 280, L11
\bibitem{} Vladilo, G., Centuri\'on, M., de Boer, K. S., King, D. L.,
Lipman, K., Stegert, J. S. W., Unger, S. W., \& Walton, N. A. 1994,
\aap, 291, 425
\bibitem{} Vogt, S. S. 1992, in Proc. ESO Workshop 40, High Resolution
Spectroscopy with the VLT, ed. M.-H. Ulrich (Garching: ESO), 223
\bibitem{} Vogt, S. S. 1994, UCO/Lick Observatory Technical Report
No. 76
\bibitem{} Wade, R. A., \& Horne, K. D. 1988, \apj, 324, 411
\bibitem{} Wang, L., Howell, D. A., H\"oflich, P., \& Wheeler,
J. C. 2000, \apj, submitted (astro-ph/9912033)
\bibitem{} Wang, L., \& Hu, J. 1994, \nat, 369, 380
\bibitem{} Weiler, K. W., Panagia, N., \& Sramek, R. A. 1990, \apj,
364, 611
\bibitem{} Weiler, K. W., Sramek, R. A., Panagia, N., van der Hulst,
J. M., \& Salvati, M. 1986, \apj, 301, 790
\bibitem{} Wheeler, J. C., et al. 1993, \apj, 417, L71
\bibitem{} Wheeler, J. C., \& Filippenko, A. V. 1996, in Supernovae and
Supernova Remnants, ed. R. A. McCray \& Z. Wang (Cambridge: Cambridge
University Press), 241
\bibitem{} Wild, P. 1960, \pasp, 72, 97
\bibitem{} Williams, R. 1987, \apj, 320, L117
\bibitem{} Woosley, S. E., Eastman, R. G., Weaver, T. A., \& Pinto,
P. A. 1994, \apj, 429, 300
\bibitem{} Woosley, S. E., Pinto, P. A., Martin, P. G., \& Weaver,
T. A. 1987, \apj, 318, 664
\bibitem{} Young, T. R., Baron, E., \& Branch, D. 1995, \apj, 449, L51
\bibitem{} Zhang, Q., \& Wang, Z. R. 1996, \aap, 307, 166
\bibitem{} Zimmerman, H.-U., et al. 1994, \nat, 367, 621


\end{thebibliography}
\end{document}


\clearpage
\ptlandscape
\clearpage
\pagestyle{empty}
\begin{deluxetable}{lccccccccccccl}
\textheight=5.5in
\clearpage
\tablenum{1}
\tablewidth{680pt}
\tablecaption{JOURNAL OF OBSERVATIONS\label{tbl-1}}
\tablehead{\colhead{Day\tablenotemark{a}} & 
\colhead{UT Date} &
\colhead{Tel.\tablenotemark{b}} &
\colhead{Range\tablenotemark{c}}  &
\colhead{Res.\tablenotemark{d}} &
\colhead{P.A.\tablenotemark{e}} &
\colhead{Opt. P.A.\tablenotemark{f}} &
\colhead{Air.\tablenotemark{g}} & 
\colhead{Flux Std.\tablenotemark{h}} &
\colhead{Ph.\tablenotemark{i}} &
\colhead{See.\tablenotemark{j}} &
\colhead{Slit} &
\colhead{Exp.\tablenotemark{k}} &
\colhead{Observer(s)} \\
\colhead{} &
\colhead{} &
\colhead{} &
\colhead{(\AA)} &
\colhead{(\AA)} &
\colhead{($^\circ$)} &
\colhead{($^\circ$)} &
\colhead{} &
\colhead{} &
\colhead{} &
\colhead{($^{\prime\prime}$)} &
\colhead{($^{\prime\prime}$)} &
\colhead{(s)} & 
\colhead{} }
\startdata
3   & 1993-03-30 & L  & 3560-7300 &  6    & 137    & 136   & 1.2   & F34         &   N      &  1.5 & 2 &  180             & MD, DS           \nl
16  & 1993-04-12 & L  & 3340-7460 &  6    & 25     &20-24  & 1.2   & F34/HD84    &   N,S    & 2.0 & 2 &  100-300        & DSF                  \nl
17  & 1993-04-13 & L  & 3120-10300&  6-15 &152-188 &151-192& 1.2   & F34/HD84    &   P,S    &  1.2 & 2 &  50-300          & AVF, TM     \nl
18  & 1993-04-14 & L  & 3120-10300&  6-15 &132-156 &133-159& 1.2   & F34/HD84    &   P,S    &  1.5 & 2 &  40-300          & AVF, TM     \nl
19  & 1993-04-15 & L  & 3120-10300&  6-15 &111-188 &114-194& 1.2   & F34/BD26    &   P,S    &  2.0 & 2 &  80-800          & AVF, TM     \nl
32  & 1993-04-28 & L  & 3200-7380 &  3-6  & 177    & 177   & 1.2   & F34/BD26    &    Y     & 1.5 & 2 &  300              & SLH                  \nl
33  & 1993-04-29 & L  & 3200-7380 &  3-6  & 177    & 178   & 1.2   & F34/BD26    &    Y     & 1.5 & 2 &  300              & SLH                  \nl
34  & 1993-04-30 & L  & 3120-9900 &  6-7  & 89     & 87-95 & 1.5   & F34/BD26    &   Y,S    &  1.5 & 2 & 50-400           & AVF, LCH     \nl
45  & 1993-05-11 & L  & 3120-9900 &  6-7  &117-130 &118-132& 1.2   & F34/BD26    &   N,S    &  3.5 & 2 &  100-900         & AVF, TM     \nl
51  & 1993-05-17 & L  & 4221-5580 &  3    & 145    & 147   & 1.2   & F34         &   Y      &  2.0 & 2 &  300-600       & DF                   \nl
56  & 1993-05-22 & L  & 3250-10400& 6-15  & 135    & 127   & 1.3   & F34         &   P,S    &  2.5 & 2 &  200             & HS, RAG               \nl
83  & 1993-06-18 & L  & 3220-10400& 6-15  & 90     & 98    & 1.5   & BD28/BD26   &   N      & 2.0 & 2 &  300              & HES, CJL    \nl
89  & 1993-06-24 & L  & 3340-8050 &  6-7  & 90     & 98    & 1.5   & F56         &   N      & 2.0 & 2 &  60-300           & RHB, RLW     \nl
91  & 1993-06-26 & L  & 3480-8150 &  6-7  & 90     & 93    & 1.5   & F34         &   N      & 2.0 & 2  &  300             & MAB, DCK, GDW        \nl
93  & 1993-06-28 & L  & 3110-10400&  6-15 & 84     & 82-86 & 1.6   & F34/BD26    &   P,S    &  2.0 & 1.8 &  200-400       & AVF, TM       \nl
109 & 1993-07-14 & L  & 3110-10400&  6-15 & 18     & 5-14  & 3.5   &F34/BD26/HD19&   Y,S    &  3.5 & 2 & 200-600          & AVF, TM, AJB        \nl
123 & 1993-07-28 & L  & 3350-10350&  6-15 & 72-147 &149-155& 2.9   &BD28/BD26    &   Y,S    &  2.0 & 2 & 50-400           & AVF, TM, AJB        \nl
139 & 1993-08-13 & L  & 3160-9880 &  6-7  &153-158 &149-160& 3.0   &BD28/BD26    &   Y,S    &  2.5 & 2 & 100-600          & AVF, TM     \nl
167 & 1993-09-10 & L  & 3180-9910 &  6-7  &118-130 &116-131& 2.2   & BD28/BD26   &   Y,S    &  1.5 & 2 & 150-900          & AVF, TM     \nl
182 & 1993-09-25 & L  & 3120-9920 &  6-7  & 91-95  & 89-95 & 1.5   & F34/HD19    &   Y,S    &  2.5 & 2 & 500-600          & AVF, TM     \nl
209 & 1993-10-22 & L  & 3130-9910 &  6-7  & 75-85  & 75-87 & 1.4   & F34/HD84    &   N,S    &  1.0 & 2 & 100-900          & AVF, TM     \nl
226 & 1993-11-08 & L  & 3120-9920 &  6-7  & 213-223&212-225& 1.2   & BD28/BD17   &   N      &  1.5 & 2 & 600-900          & AVF, TM     \nl
235 & 1993-11-17 & L  & 3120-10300&  6-15 & 61-81  & 62-83 & 1.4   & F34/BD17    &   N      &  1.0 & 2 & 900-1500         & AVF, TM      \nl
266 & 1993-12-18 & L  & 3120-9910 &  6-7  &218-252 &219-268& 1.3   & F34/BD17    &   N      &  2.0 & 2 & 1200-1800        & AVF, LCH             \nl
286 & 1994-01-07 & L  & 3260-9900 &  6-7  &138-155 &139-157& 1.2   &F34/BD17/BD26&   N      &  1.5 & 2 & 900-1800         & AVF, TM     \nl
298 & 1994-01-19 & L  & 3120-9910 &  6-7  &214-233 &216-234& 1.2   &F34/BD17/BD26&   N      &  1.0 & 2 & 1200-1800        & AVF, TM      \nl
315 & 1994-02-05 & L  & 3120-9920 &  6-7  &90-116  &91-117 & 1.4   & F34/HD19    &   N      &  1.5 & 2 & 900-1800         & AVF                  \nl
355 & 1994-03-17 & L  & 3120-8020 &  6-7  &118-144 &111-144& 1.2   & F34/BD26    &   N      &  2.0 & 2 & 750-1800         & AVF, TM, AJB        \nl
\tablebreak
387 & 1994-04-18 & L  & 3120-9900 &  6-7  &277-299 &278-299& 1.3   & F34/HD84    &   N      &  1.5 & 2 & 1800-2700        & AVF, LCH     \nl
433 & 1994-06-03 & L  & 3120-12000&  6-15 &81-95   &81-93  & 1.5   & F34/HD84    &   N      &  2.0 & 2 & 1200-1500        & AVF, TM, AJB         \nl
473 & 1994-07-13 & L  & 3120-11400&  6-15 & 17     &-3.5-11& 3.4   & BD28/BD17   &   P      &  2.0 & 2 &  1800            & AVF           \nl
523 & 1994-09-01 & L  & 3120-8030 &  6-7  &136-169 &134-165& 3.0   & BD28/BD26   &   Y      &  4.0 & 2 &  1800            & AVF, TM, AJB         \nl
525 & 1994-09-03 & L  & 3120-8030 &  6    &126-134 &125-133& 2.3   & BD28/BD26   &   Y      &  2.5 & 2 &  1800            & AVF, TM, AJB  \nl
553 & 1994-10-01 & L  & 3120-8040 &  6-7  & 129    &116-126& 1.2   & BD28/BD26   &   Y      &  1.5 & 2 &  1800            & TM, LCH      \nl
670 & 1995-01-26 & K-I & 3880-8800& 2.5-9 & 180    &170-188& 1.5   & BD26        &  Y       &  1.3 &     1& 300-450       & AVF, LCH     \nl
    &            &     & 5628-6939& 2.5   & 180    &  182  & 1.5   & BD26        &  Y       &  1.3 & 0.7  &     900       & AVF, LCH     \nl
881 & 1995-08-25 & L  & 4260-7010 &  7    & 197    & 132   & 2.4   & BD26        &  Y       &  2.0 & 4 &  300-1800        & AVF, DCL     \nl
976 & 1995-11-28 & K-I & 3930-8970& 9     & 47     &    47 & 1.7   & BD17        &  N       &  1.3 &    1& 300-600        & AVF, AJB    \nl
    &            &     & 5784-7092& 2.5   & 47     & 39    & 1.7   &      HD84   &  N       &  1.3 &0.7  &     600        & AVF, AJB    \nl
1766& 1998-01-26 & K-II& 3920-8950& 5-8   & 150    &149-155& 1.6   & BD26        &  Y       &  1.0 & 1 & 300-900          & AVF, DCL     \nl
2028& 1998-10-15 & L  & 3300-10200&  6-15 & 250    & 253   & 1.4   & F34/HD84    & N        &  2.5 & 2 &  1800            & AVF, DCL     \nl
2069& 1998-11-25 & L  & 3340-10200&  6-15 & 68     & 52    & 1.3   & BD28/BD17   &  N       &  2.5 & 2 &             2700 & AVF, TM      \nl
2115& 1999-01-10 & L  & 3350-10300&  6-15 & 180    & 175   & 1.2   & BD28/BD17   & N        &  1.5 & 2 &             1800 & AVF, TM, WL  \nl
2176& 1999-03-12 & L  & 3280-10500&  6-15 & 186    & 157   & 1.2   & F34/HD19    & N        &  2.0 & 2 &             1800 & AVF, TM, WL  \nl
2454& 1999-12-15 & K-II& 3900-8900&  9    & 205    & 207   & 1.6   & BD17        &  Y       &  1.1 & 1 &             600  & AVF          \nl

\enddata
\tablecomments{For each day, multiple observations are combined into a one-line summary.  For days 670 and 976,
the exposures with different slit widths are listed separately.}
\tablenotetext{a}{Days since estimated date of explosion, 1993 March 27 UT (JD 2,449,074).  Note that the data from days
523 and 525 are combined into one spectrum in the figures.}

\tablenotetext{b}{L = Lick 3-m/Kast Double Spectrograph; K-I =
Keck-I 10-m/Low Resolution Imaging Spectrometer (LRIS); K-II = Keck-II 10-m/LRIS.}

\tablenotetext{c}{Observed wavelength range of spectrum.  In some cases, the
extreme ends are very noisy, and are not shown in the figures.}

\tablenotetext{d}{Approximate spectral resolution (full width at half maximum intensity).}

\tablenotetext{e}{Position angle of the spectrograph slit.  Most
observations taken within 10$^\circ$ of the parallactic angle
(Filippenko 1982).  On day 123,
difficult conditions arising from observing under the pole with
an equatorially mounted telescope required constant rotation of the slit
to observe the supernova; the orientation was within 10$^\circ$ of
the parallactic angle for the duration of the observation.  For day 881, a specific
position angle (197) going through a nearby star was used to ensure that SN 1993J was in the slit.  On days 83 and 
89, the position angle used to observe the standard stars was significantly different
from the optimal parallactic angle, but the airmass of the observations was low.}

\tablenotetext{f}{Optimal parallactic angle over the course of the exposures.}

\tablenotetext{g}{Average airmass of observations.}

\tablenotetext{h}{The standard stars are as follows: F34 = Feige~34, F56 = Feige~56, BD28 = BD+28$^\circ$4211---Stone (1977),
 Massey \& Gronwall (1990);
HD19 = HD~19445, HD84 = HD~84937, BD26 = BD+26$^\circ$2606, BD17 = BD+17$^\circ$4711---Oke \& Gunn (1983).}

\tablenotetext{i}{Estimation of the photometric conditions for the night:  N = not photometric, P = possibly
photometric, Y = photometric, S = 2$\arcsec$ slit width observations scaled to 8$\arcsec$ slit width observations.}

\tablenotetext{j}{Seeing is estimated from the data and observers' records.}

\tablenotetext{k}{The range of exposure times for the various observations.  In some cases, multiple 
exposures were taken so the actual time of integration at any given wavelength may be $3-4$ times the 
longest exposure time listed.}

\end{deluxetable}

\clearpage